\documentclass[twocolumn,aps,showpacs,groupedaddress]{revtex4-1}
\usepackage{array}
\usepackage{graphicx}
\def\be{\begin{equation}}
\def\ee{\end{equation}}
\def\bea{\begin{eqnarray}}
\def\eea{\end{eqnarray}}
\def\ba{\begin{array}}
\def\ea{\end{array}}
\def\bdm{\begin{displaymath}}
\def\edm{\end{displaymath}}

\begin{document}

\title{Preparation of two-particle total hyperfine spin singlet states via spin-changing dynamics}

\author{Chao-Chun Huang$^1$, M.-S. Chang$^2$ and S.-K. Yip$^{1,2}$}

\affiliation{Institute of Physics$^1$, Institute of Atomic and Molecular Sciences$^2$,
 Academia Sinica, Taipei, Taiwan}

%\date{\today }

\begin{abstract}

We present our proposals for generating total hyperfine spin zero state
for two $f=1$ or two $f=2$ particles, starting from initial unentangled states.
We show that our goal can be achieved by exploiting spin changing dynamics
and quadratic Zeeman shifts
with realistic choices of external magnetic fields and evolution time intervals.

\end{abstract}


% Atoms in optical lattices
%     ultracold gases,
% 75.10.Jm 	Quantized spin models, including quantum spin frustration

\maketitle

\section{Introduction}

State preparation for many-particle systems is gaining increasing attention recently. States with desired quantum entanglement properties find important applications in quantum information
processing, cryptography, and computation \cite{Bennett}.  Indeed efforts
in this direction can be found in areas as diverse as quantum  dots \cite{Hanson07},  trapped ions \cite{Kim09}, and cold atoms \cite{Anderlini07,Trotzky10}, to name a few.  It has also been recognized
recently that similar ideas can be used to create exotic many-body states
in cold atoms. Many of these states are of particular interest in the condensed matter context.  Examples of such proposals include the Haldane state for a spin-1 lattice \cite{Garcia04},  resonant valence bond states \cite{Trebst06},
antiferromagnetic states \cite{Rey07,Barthel09},
d-wave ``superconductivity'' \cite{Trebst06,Rey09},
Luttinger liquids \cite{Rahmani11}, and even states that correspond to highest energy (instead of ground) state
for a given Hamiltonian \cite{Garcia04,Sorenson10}.
%This list is certainly not complete.

In this paper, we would like to discuss our proposal to generate a total hyperfine spin-zero singlet pair, starting from two particles that each has hyperfine spin-1 or 2. Our primary motivation is that
the ground states of certain one (spatial) dimensional lattice of Bosons
with one atom per site and with suitable interaction between the Bosons
are expected to be in the dimerized state \cite{dimer,Pochung11}.
In this state, the lattice translational symmetry is spontaneously
broken so that the spin-correlation between neighboring sites
alternate between strong and weak.  Furthermore,  it
can be shown that this dimerized state is smoothly connected
to the ground state of a new Hamiltonian where alternate bonds
are weakened \cite{Pochung11,Kitazawa99}.  In the limit where the alternate bonds
are weakened completely to zero, we have a collection of
isolated singlet pairs.  The system remains gapped
throughout this process.  Reversing the argument, once
singlet pairs can be prepared, in principle one can then
obtain the many-body dimerized state of the original Hamiltonian
by gradually turning on the coupling between the pairs.
A similar argument can also be applied generally to valence bond states.
Thus preparing the singlets would be a useful first step
in generating these unconventional magnetic states.

Irrespective of whether we can ultimately obtain the dimerized
states experimentally using this method, we believe that
our proposals of generating singlet pairs of hyperfine spin-1 and spin-2
particles are of interest in their own right.
While most literature on quantum information and cryptography deals
with particles with only two internal states (e.g. spin-up versus down for spin-$1/2$ particles or horizontal versus vertical polarization for photons), it is known that higher dimensional quantum systems ({\it i.e.}
systems with more than two internal states) offer many advantages such as higher tolerance in noise in quantum key distribution \cite{QKD} or tests of Bell inequalities \cite{Bell}. A spin-singlet state of two spin-1 (2) particles is (one of) the most entangled two-particle state in a dimension 3 (5) quantum system. It is therefore of high interest to know how to generate them, and here we offer a scheme starting from completely unentangled initial states\cite{expr}.

We shall discuss two schemes. The first is simply to vary an external magnetic
field adiabatically.  This scheme is simple, but would work only in very
special cases.  The second, the central work of this paper, is a dynamical
scheme. The main idea is to exploit the spin changing dynamics of two interacting particles in the presence of a (generally time-dependent) quadratic Zeeman field.  This dynamics has been studied experimentally by Widera et al \cite{Widera06}.  However, their focus was to extract the spin dependent interaction parameters (the spin dependent s-wave scattering lengths).  We shall show that,
for proper choice of parameters, one can prepare pure spin-singlet pairs using
this spin changing dynamics.

We introduce our proposal for hyperfine spin-1 particles in section \ref{sec:one}. The case for hyperfine spin-2 particles
is discussed in section \ref{sec:two}.
In these two sections, we emphasize the basic ideas and simplify our analysis by putting the magnetic field to be zero during some time intervals.  The generalization to the cases where
the magnetic fields are always finite
%to avoid possible undesirable effects of zero field
are discussed in section \ref{sec:three}.
The conclusions are in section \ref{sec:four}. In appendix A
we provide some details of our estimate on the fidelity of our scheme.
%we show how
%to estimate influence of inaccuracies in the time intervals used in evolution
%which may arise in experiments.
For simplicity, we shall occasionally use "spins" for hyperfine spins when there is no danger of confusion.

\section{HYPERFINE SPIN-1 SYSTEMS}\label{sec:one}

We assume that one can initially prepare two particles in a potential well
(or a double well, the detailed shape  of  the potential does not matter),
each with hyperfine state $ | f, m_f\rangle = | f, 0\rangle $.   In this (next) section
$f = 1$ ($f = 2$).  This initial state has also been obtained by Widera
et al \cite{Widera06,note}.  We shall assume that these particles
reside and stay in the lowest motional  ground state throughout the dynamics, thus their spins
are the only degrees of freedom.  A quantum mechanical state can therefore
be indicated only by the quantum numbers $m_f$'s of the two particles.  For simplicity
we shall suppress the $f$ labels.  Thus $|0, 0\rangle _m$ would indicate that
both particles have $m_f = 0$, whereas
$\frac{1}{\sqrt{2}}  \left( | 1, -1\rangle _m + | -1, 1\rangle _m \right)$ indicates
that there is one particle each in the $m_f = \pm 1$ state.
Since  we are dealing with Bosons, only symmetric states need to be kept.
The subscripts $m$ remind us that we are using the $m_f$ basis.

Since we are interested in obtaining a total spin singlet, we would
also like to keep track of the total hyperfine spin $F$ of the two particles.
It is therefore also convenient to use the basis $ | F M_F\rangle _F $ where
$M_F$ is the projection of $F$.  We only need to consider even $F$'s.
The subscripts $F$  would be used to remind us when we are using this basis.
The conservation of total $m$ implies that we only need to
consider the two states $|0 0\rangle _F$ and $| 2 0 \rangle _F$.

We consider the time evolution of these particles under the presence of
a (in general time-dependent) quadratic Zeeman field.  The linear
Zeeman field needs not be considered due to the conservation of total $m$
and hence $M$ within our proposed schemes.

The Hamiltonian consists of simply two  terms:  the quadratic Zeeman energy
and the interaction energy.  The first contribution is a single particle term.
It represents a $m_f$ dependent energy quadratic in magnetic field $B$ (for not too
large $B$'s) and can
be written in the form
$H_Q = q \left( \sum_{m_1,m_2} ( m_1^2 + m_2^2) | m_1, m_2\rangle \langle m_1, m_2 | \right)$
where $q = \hat q B^2$ with $\hat q$ a coefficient depending on the atom under consideration.
The interaction energy is diagonal in the total hyperfine spin basis
$|0 0 \rangle _F$ and $| 2 0 \rangle _F$.
We shall denote these interaction energies
by $E_0$ and $E_2$ respectively.

We first discuss the adiabatic scheme.  The singlet state can be generated
by adiabatically varying the external field $B$ hence $q$ if both
the initial and final states are either the highest or the lowest energy states
when $q$ varies.  Suppose $q$ is positive.  Then the initial state
$ | 0, 0\rangle _m$ is the lowest energy state for large $q$.  If $E_2 > E_0$,
then the singlet state is also the ground state at $q = 0$.  Hence,  if
the initial $ | 0, 0\rangle _m$ state was produced at sufficiently large $q$, then
decreasing the magnetic field to zero would automatically generate the
singlet $ |0 0\rangle _F$ state.  Such a scheme would not work of $E _2 < E_0$ \cite{qsign},
or their difference is too small for adiabaticity to be satisfied
for $q$ to be varied in time.

Below we turn to our dynamical scheme which works for both $E_2 > (<) E_0$.
Let us first set up the Hamiltonian. In the
$ | 0 0 \rangle _F$ and $|2 0\rangle _F$ basis,  the interaction part
of the energy is simply a diagonal matrix,
\bdm
H_I = \left(
\ba{cc}
E_0 & 0 \\
0 & E_2
\ea
\right).
\edm
Since we are targeting  the state $|0 0\rangle_F$ as the final state, we shall work in
this basis.  It is simple to rewrite $H_Q$ in this $|F M_F\rangle$ basis since,
from the Clebsch-Gordan coefficients,  we know that
\bea
| 0 0 \rangle _F &=&   \frac{1}{\sqrt{3}} \left( | 1, -1\rangle _m   + | -1, 1\rangle _m - |0, 0\rangle _m\right)
\nonumber \\
 | 2 0\rangle _F &=&   \frac{1}{\sqrt{6}} \left( | 1, -1\rangle _m   + | -1, 1\rangle _m  + 2 |0, 0\rangle _m\right),
 \label{transf1}
 \eea
 from which we find
 \bdm
 H_Q = \frac{2q}{3} \left(
 \ba{cc}
 2 & \sqrt{2} \\
 \sqrt{2} & 1
 \ea
 \right).
  \edm
  Hence the total Hamiltonian is
  %in the $|0 0>_F$ and $ |2 0>_F$ basis,
 \bea
 H = \left(
 \ba{cc}
 E_0 & 0 \\
 0 & E_2
 \ea
 \right)
 +
  \frac{2q}{3} \left(
 \ba{cc}
 2 & \sqrt{2} \\
 \sqrt{2} & 1
 \ea
 \right).
 \label{H1}
 \eea
 In this basis, our initial state $|0,0\rangle _m$ is simply
 $(u(0),v(0))^T = (- 1/\sqrt{3}, \sqrt{2/3})^T$  where  $T$ denotes the transpose.

 Let us consider the evolution of the state vector
  $(u(t),v(t))^T$ in time $t$.  Its value can be easily found
 since we have simply a two-level system.  $H$ can be rewritten
 as $ \left( \frac{E_0+E_2}{2} + q \right) + \vec H_{eff} \cdot \vec \tau$
 where $\vec \tau$ are the Pauli matrices and
 $H_{eff,x} = \frac{2 \sqrt{2} q}{3}$ and
 $H_{eff,z} =  \frac{E_0- E_2}{2} + \frac{q}{3}$.  The scalar term
 only gives an overall phase and will be dropped below.  The state
 vector at time $t$ is then given by
 \bea
 \left( \ba{c} u(t) \\ v(t) \ea \right)
 =
 \left( {\rm cos} \Omega t - i
 \frac{ \vec H_{eff} \cdot \vec \tau}{\Omega} {\rm sin} \Omega t \right)
  \left( \ba{c} u(0) \\ v(0) \ea \right),
  \eea
  where
  \be
  \Omega \equiv |\vec H_{eff} | =
  \left[ \left(  \frac{E_0- E_2}{2} + \frac{q}{3} \right)^2 + \frac{8}{9} q^2 \right]^{1/2}.   \
  \label{def-Omega}
  \ee
Here $2 \Omega$ is the Rabi frequency.
 The condition that we have a singlet at time $t = t^*$ is $v(t^*)=0$. Given that $u(0)=-1/\sqrt{3}$ and $v(0)=\sqrt{2/3}$ we need

   \be
   \left( \frac{ i H_{eff,x}}{\Omega} {\rm sin} \Omega t^* \right)
   + \sqrt{2} \left( {\rm cos} \Omega t^* +
   \frac{ i H_{eff,z}}{\Omega} {\rm sin} \Omega t^* \right) = 0, \label{int}
   \ee
   and hence

  \be
  {\rm cos} \Omega t^* = 0   \label{condt}
  \ee
  and
  \be
  H_{eff,x} + \sqrt{2} H_{eff,z} = 0. \label{condH}
  \ee
  These require, respectively,
  \be
  t^* = \left( n + \frac{1}{2} \right) \pi / \Omega
  \ee
  and
  \be
  q =  \frac{E_2- E_0}{2}.  \label{condq}
  \ee
  This last expression and Eq. (\ref{def-Omega}) together give
  \be
  \Omega =    \frac{|E_2- E_0|}{\sqrt{3}}. \label{Omega}
  \ee
  Hence if $q$ (hence $\hat q$) has the same sign as $E_2 - E_0$, we
  obtain a singlet at times $ t^* = \frac{\pi}{2 \Omega}, \frac{3\pi}{2 \Omega}$ ...
  As an example, consider $^{23}$Na atoms in its $f=1$ hyperfine
  state, where $\hat q > 0$.  The s-wave scattering lengths
  satisfy $a_2 > a_0$, hence $E_2 > E_0$.  Hence it is possible
  to obtain a pure singlet at time
  $t^*$  when we choose the magnetic
  field to be $B = \left(   \frac{E_2- E_0}{2 \hat q}  \right)^{1/2}$.
  The shortest possible time  is
  $   t^* = \frac{\pi}{2 \Omega} =      \frac{\sqrt{3} \pi}{2 (E_2 - E_0)}$.
  If the spatial wavefunction is $\psi(\vec r)$, then $E_F = \frac{ 4 \pi a_F}{M_a} \int d^3 \vec r | \psi (\vec r)|^4 = a_F \tilde U$, where $\tilde U$ is defined by this expression as in \cite{Widera06}, and $M_a$ is the mass of the atoms. Immediately following that $E_2 - E_0 = \tilde U (a_2 - a_0) $, and for $^{23}$Na $a_2 - a_0$ is approximately $3.5 a_B$ \cite{Stenger98}, where $a_B$ is the Bohr radius. Given that $\hat q = 278$ Hz/G$^2$ and if we choose $\tilde U a_B = (2 \pi) 30$ Hz, then the required magnetic field and the time to obtain a singlet are calculated to be $B=1.089$ G and $t=4.1$ ms, respectively.

As mentioned above this dynamical scheme is directly applicable to $^{23}$Na, where $\hat q$ and $E_2 - E_0$ are of the same sign. With simple modifications, this scheme can be generalized to cases where $\hat q$ and $E_2 - E_0$ are of the opposite sign, such as $^{87}$Rb. This is done by simply letting the state first evolves under zero field till time $t_1$, so that the wavefunction at $t_1$ is

    \bdm
    \left( \ba{c}
    -\frac{1}{\sqrt{3}} e^{- i E_0 t_1} \\
    \sqrt{\frac{2}{3}} e^{ -i E_2 t_1} \ea \right)
    = \frac{ e^{ -i E_2 t_1}}{\sqrt{3}}
    \left( \ba{c} - e^{i \phi} \\
    \sqrt{2} \ea \right),
    \edm
    where $\phi = (E_2 - E_0) t_1$ is the phase
    difference between the $F=0$ and $F=2$ states due to the interparticle interaction during the time  interval $ 0 < t < t_1$. Then a magnetic field and hence the quadratic Zeeman energy is turned on at $t_1$, and the condition for obtaining a singlet after a further time $t^*$ (thus at total time $t = t_1 + t^*$) is simply the same as Eq.~(\ref{int}), except an extra factor $e^{i \phi}$ multiplying the first term. In this case condition (\ref{condH}) is now replaced by
    \be
    H_{eff,x} {\rm cos} \phi + \sqrt{2} H_{eff,z} = 0, \label{condH1}
    \ee
    which requires
    \be
    q \left[ \frac{ 1 + 2 {\rm cos} \phi}{3} \right]
    =  \frac{E_2- E_0}{2}. \label{condq1}
    \ee
With suitable choice of $\phi$ (\textit{i.e.,} $t_1$), this condition can be satisfied even if $q$ and $E_2 - E_0$ are of opposite signs. For this $q$, the frequency $\Omega$ is given by
    \be
    \Omega = \frac{|E_2 - E_0|}{ | 1 + 2 {\rm cos} \phi | }
      \left( 2 + {\rm cos}^2 \phi  \right)^{1/2}, \label{condq2}
    \ee
    and the required time $t^*$ is given by
    \bea
    {\rm cot} \Omega t^* &=& \frac{H_{eff,x}}{\sqrt{2} \Omega} {\rm sin} \phi   \nonumber \\
    &=& ({\rm sgn} q ) \frac{ {\rm sin} \phi}     {\left( 2 + {\rm cos}^2 \phi  \right)^{1/2}}. \label{condq3}
    \eea

\begin{figure}
\vspace{-0.0cm}
\includegraphics[width=0.49\textwidth]{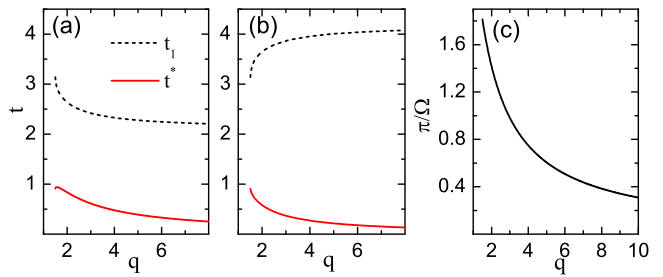}
\vspace{-0.6cm} \caption {Frames (a) and (b):  two sets of solutions for
$t_1$ and $t^*$ as functions of $q$.
In frame (c), the value of $ \pi / \Omega $ is shown.
Energy and time are in units of $E_0-E_2$ and $1/(E_0-E_2)$, respectively.}
\label{fig1}
\end{figure}

It follows that for a solution $(\phi(t_1), t^*)$ solved by Eq.~(\ref{condq1}-\ref{condq3}), $(\phi+n2\pi , t^*+ m\pi /\Omega )$ and $(-\phi+n2\pi, -t^*+ m\pi /\Omega)$ are also solutions, where $n$ and $m$ are integers, and $\Omega$ is an implicit function of $\phi$.
    Of the solutions above only positive $t_1$ and $t^*$'s are realistic, and Eq.~(\ref{condq1}) implies that there is no solution when $-1.5< \frac{q}{E_2-E_0}<0.5$ which holds for arbitrary signs of $q$ and $E_2 - E_0$.
   Fig.~\ref{fig1}(a)(b) show two possible sets of $(t_1,t^*)$, where $E_2<E_0$ and $q>0$.
   In Fig.~\ref{fig1}(c) the value of $\pi / \Omega$ is shown.
   It is note that the solutions in Fig.~\ref{fig1} are also applicable
   for cases where $q<0$ and $E_2>E_0$,
   due to the symmetry properties of equations~(\ref{condq1}) and (\ref{condq3}).
     When the signs of $q$ and $E_2 - E_0$ are both reversed, it is observed that $H \to - H$, up to a constant, and $\phi \to - \phi$ in Eq.~(\ref{condq1}).

   Consider $^{87}$Rb atoms in its $f=1$ hyperfine state,
   where the value of $a_2-a_0$ is approximately $-1.4 a_B$ \cite{Widera06}.
   By choosing $\tilde U a_B = (2 \pi) 30$  Hz, then $E_0-E_2 = 264$Hz. If we further choose $q=1.5(E_0-E_2)$, then the required magnetic field is $B=2.35$ G.
   Here we have used $\hat q = 71.7$ Hz/G$^2$ for $^{87}$Rb.
   For this particular choice of $q$ the solutions
   in Fig.~\ref{fig1}(a) and (b) are same and for the times
   are $t_1=11.90$ ms and $t^*=3.435$ ms.

   We shall estimate the error in producing the singlet state due to inaccuracies  $\Delta t_1$ and $\Delta t^*$ for $t_1$ and $t^*$, respectively. Let $t_1$ and $t^*$ be the times needed to generate the singlet state so that $|u(t)|=1$ and $|v(t)|=0$, then the expected value of $|u(\Delta t_1,\Delta t^*)| = \sqrt{1-|v(\Delta t_1,\Delta t^*)|^2}\approx1-\frac{|v|^2}{2}$.
Here the approximation holds as long as $\Delta t_1$ and $\Delta t^*$ are not too large, and the analytical expression of $|v(\Delta t_1,\Delta t^*)|$ is shown in Appendix~\ref{app1}. Take for instance, if $|\Delta t_1|+|\Delta t^*| < 0.31$ ms for $^{87}$Rb, then $|v|^2 < 0.02$ and $|u| > 0.99$.    Similarly for the $^{23}$Na, if the total inaccuracies of time $|\Delta t_1|+|\Delta t^*| < 0.36$ ms, the same fidelity is achieved. We also note that from Appendix~\ref{app1} and Fig.~\ref{fig1}(c), the value of $|u(t)| $ is closer to 1 for the same inaccuracies in time if a smaller value of $|q|$ is used.

   In the above, we have illustrated our scheme where zero magnetic field is utilized. In real experiments, zero magnetic field is not easily attainable, and undesirable effects may occur in low fields. To avoid those practical issues, our scheme can be readily generalized to the case of finite field, by adjusting slightly the second applied field, provided that the first field is not too large. Before we address this issue in section \ref{sec:three}, we shall discuss generalization of this dynamical scheme to $f=2$ systems.

\section{HYPERFINE SPIN-2 SYSTEMS}      \label{sec:two}

The considerations here are direct generalization of those in the last section, and let us first consider the adiabatic scheme.  Suppose $q < 0$, as in the case
of $f = 2$ hyperfine states of $^{23}$Na and $^{87}$Rb, then the initial
state $|0,0>_m$ is the highest energy level for large magnetic fields.
In cases where $E_0 > E_{2,4}$, then adiabatically switching off the external field would generate the desired singlet state. Conversely, if $q > 0$,
the singlet state can be obtained if $E_{2,4} > E_0$.  This scheme however
is limited in its applicability, and in particular it neither works for $^{23}$Na nor $^{87}$Rb, where $q < 0$ and
$a_{2,4} > a_0$.

Let us proceed to consider the dynamical scheme. Similar to $f=1$ spinors, we can write down the Hamiltonian for $f=2$ spinors in the $|0 0\rangle_F$, $ |2 0\rangle_F$ and $ |4 0\rangle_F$ basis, where
\bea
&| 0 0\rangle _F = \frac{1} {\sqrt{5}}  [ |0, 0\rangle  - ( |1,-1\rangle  + |-1,1\rangle ) \nonumber\\
&\texttt{\ \ \ \ \ \ }+(|2,-2\rangle  +|-2,2\rangle ) ]_m,     \nonumber \\
&| 2 0 \rangle _F = \frac{1}{\sqrt{14}} [ -2|0, 0\rangle  + ( |1,-1\rangle  + |-1,1\rangle ) \nonumber\\
& \texttt{\ \ \ \ \ \ }   + 2(|2,-2\rangle  +|-2,2\rangle ) ]_m,   \\
&| 4 0 \rangle _F = \frac{1}{\sqrt{35}} [ \sqrt{18} |0, 0\rangle  + \sqrt{8} ( |1,-1\rangle  + |-1,1\rangle )  \nonumber\\
& \texttt{\ \ \ \ \ \ }   + \frac{1}{\sqrt{2}} (|2,-2\rangle  +|-2,2\rangle ) ]_m.  \nonumber
\label{transf2}
\eea
The Hamiltonian is
 \bea
 H = \left(
 \ba{ccc}
 E_0 & 0 &0\\
 0 & E_2 &0\\
 0 & 0 & E_4
 \ea
 \right)
 +
 {2q} \left(
 \ba{ccc}
 2 & \frac{14}{\sqrt{70}} & 0 \\
 \frac{14}{\sqrt{70}} & \frac{17}{7} & \frac{12}{7\sqrt{5}} \\
 0 &\frac{12}{7\sqrt{5}} &\frac{4}{7}
 \ea
 \right),
 \label{H2}
 \eea
where the two terms represent the interaction energy and the quadratic Zeeman effect, respectively. In this basis, our initial state $|0,0\rangle _m$ is simply $|\Psi_0\rangle  = (1/\sqrt{5}, -\sqrt{2/7}, \sqrt{18/35})^T$, and the desired final state is the singlet state $ |0 0\rangle _F = (1, 0, 0)^T$.
In other words with initial state $|\Psi_0\rangle $, we want to find a time-evolution operator $\hat{U}$
so that
\bea
_F\langle 2 0|\hat{U}|\Psi_0\rangle = _F\langle 4 0|\hat{U}|\Psi_0\rangle =0. \label{eqnf2}
\eea
Here we need to solve for two complex equations, instead of just one ($v(t)=0$) for $f=1$ spinors. In the last section, we showed that our goal is achieved by evolving the system in two time intervals with a chosen $q$ in each interval.  Following this line, we attempt our goal with the following sequence: we first let the system evolve at $q=0$ for $t_1$, and then at $q \ne 0$ for $t_2$, followed by $q=0$ for $t_3$, then finally at $q \ne 0$ again for $t_4$. Thus the time-evolution operator is
    \bea
    \hat{U}(t_1,t_2,t_3,t_4)&=& e^{ {-i H t_4}} e^{ {-i H_0 t_3}  }
    e^{ {-i H t_2} }  e^{{-i H_0 t_1}  } \nonumber \\
    &\equiv& \hat{U}_4\hat{U}_3\hat{U}_2\hat{U}_1. \label{evolution}
    \eea
Here $H_0$ represents the Hamiltonian in Eq.~(\ref{H2}) with no quadratic Zeeman effect ($q=0$).

It is convenient to define $\varepsilon$($>0$) and $\theta$ by the relationships
$E_2-E_0\equiv\varepsilon \cos\theta$ and $E_4-E_0\equiv\varepsilon \sin\theta$.
The amplitude $\varepsilon$ thus defines our energy scale and its inverse defines the time scale of the dynamics.  The angle $\theta$ then characterizes the element under consideration.
 For each (normalized) $q$ and $\theta$, we search for the times
 $t_{1-4}$ so that Eq.~(\ref{eqnf2}) is satisfied.
We have limited ourselves to total time
$t_1+t_2+t_3+t_4$ smaller than $200$ in unit of $1/\varepsilon$,
so that the total experimental time scales remain reasonably short (see below).
\begin{figure}
\vspace{-0.0cm}
\includegraphics[width=0.48\textwidth]{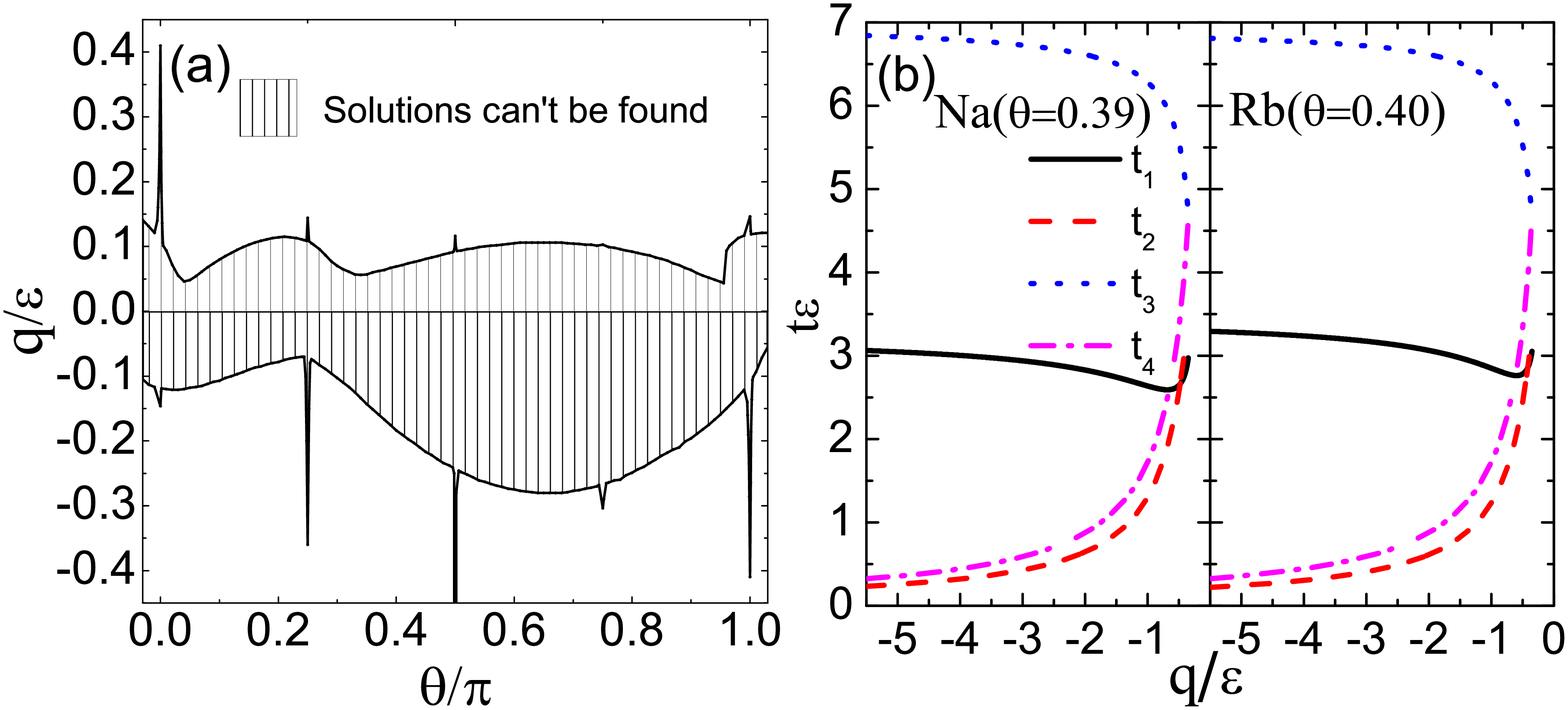}
\vspace{-0.6cm} \caption {In frame (a) the parameter regimes where we can
not find solutions are shown as shaded regions.
In frame (b), we show one solution $(t_1,t_2,t_3,t_4)$
as function of $q$
for $^{23}$Na on the left and $^{87}$Rb on the right. }
\label{fig2}
\end{figure}
There is a symmetry property:  the solutions are the same
for $q=q^*, \theta=\theta^*$ and $q=-q^*, \theta=\theta^*+\pi$
(\textit{i.e.,} $H \to - H$, up to a constant).

Fig.~\ref{fig2}(a) shows the region where no solution is found (which may be due to the limits of our computational effort).
Similar to the case of $f=1$,
when $|q|$ is small no solution is found.
Here the values of $\theta$ are limited to between $0$ and $\pi$, given the symmetry property mentioned above. We can not find a solution at $\theta=\pi/2$ $(3\pi/2)$ for $q<0$ $(>0).$ We note that for these special cases we allowed a longer search time with smaller stepsize, but no solution was found. When $q$ approaches these values, $t_{1-4}$ seem to diverge unfoundedly \cite{note1}. From the experimental point of view, if the value of $\theta$ is very close to these special cases, the system must have relatively long lifetime in order to achieve the singlet state using our scheme. However away from these values the singlet state can always
be achieved with a reasonably short time.
In Fig.~\ref{fig2}(a), there also exist sharp spikes of $q$ at $\theta=0$, $\pi/4$, $\pi/2$, $\pi$ and $3\pi/4$ \cite{note2}. For general $q$'s and $\theta$'s (excluding those discussed in the last paragraph), there exist many sets of solutions for $t_{1-4}$, and no obvious relationship is found between different sets since the time evolution is not periodic ({\it c.f.} Sec \ref{sec:one}).

Let us consider $^{23}$Na and $^{87}$Rb in the $f=2$ hyperfine state. For Na the differences of s-wave scattering lengths $(a_2-a_0,a_4-a_0)$ are approximately $(11,30)a_B$ \cite{note3}, and its quadratic Zeeman energy $q = -278\times B^2$ Hz. For $^{87}$Rb their values are $(3.5,11)a_B$ \cite{note3} and $-71.7\times B^2$ Hz. Even though $(a_2-a_0,a_4-a_0)$ are different for $^{23}$Na and $^{87}$Rb, it happens that their $\theta$'s are similar, which are approximately $0.39\pi$ and $0.40\pi$, respectively.
As solutions $t_i$'s are not unique, we only present one set of $t_{i=1-4}$ as functions of $q$ for both atoms in Fig.~\ref{fig2}(b). Take $q/\varepsilon = -0.5$, then $\varepsilon \cdot t_{i=1-4}$=(2.645, 2.571, 5.555, 3.350) for $^{23}$Na and (2.781, 2.442, 5.533, 3.343) for $^{87}$Rb. Note that the difference in $\theta$ results in
differences in $t_i$'s of up to 5\%. In real experiments if we further choose $\tilde U a_B = (2 \pi) 30$ Hz, then $\varepsilon = (2\pi) 959$ Hz for Na and $(2\pi) 346.5$ Hz for $^{87}$Rb. The controlled variables in laboratory units are $B=3.26$ G and $t_{i=1-4}=(0.450, 0.437, 0.944, 0.570)$ ms for Na and 3.90 G and $(1.307, 1.148, 2.601, 1.571)$ ms for $^{87}$Rb.

Let us also estimate the impact of inaccuracies in the time intervals used.
With our estimation in Appendix~\ref{app1},  the expected amplitude in singlet state is more than $0.99$ if the total inaccuracies of time, $|\Delta t_1|+|\Delta t_2|+|\Delta t_3|+|\Delta t_4|$,
is less than $0.011$ ms for $^{23}$Na and $0.030$ ms for $^{87}$Rb. Similar to $f=1$ cases, to increase the tolerance in the time inaccuracies, we should use smaller value of $|q|$, with a caution that when
$|q|/\varepsilon\lesssim 0.18$ for both $^{23}$Na and $^{87}$Rb, no solution may be found. In other words the $B$ must be larger than $1.95$ G ($^{23}$Na) and $2.33$ G ($^{87}$Rb) in order to arrive the singlet state.

It is also possible to find solutions by restricting to three free time intervals, such as by letting $t_1=0$, $t_2=t_4$, or $t_1=t_3$, in the expense of fixing $q$ to some special values. This is equivalent to the case in $f=1$ systems, in which $q=(E_2-E_0)/2$ when we set $t_1=0$.
\section{NON-ZERO MAGNETIC FIELD RESULTS} \label{sec:three}
In the above sections we have shown the basic ideas of our proposal, exemplified with $f=1$ and $f=2$ spinors. However, zero $B$ field environment may be difficult to prepare, and there could be undesirable effects in a zero field, such as spin-flips due to fluctuations of the magnetic field. Here we consider generalization of our scheme to the case where the magnetic field is always finite. For $f=1$, we let the system evolve at $q=q_1\neq0$ for a time $t_1$ and then at $q=q_2\neq0$ for a time $t_2$. For $f=2$, we first let the system evolve at $q=q_1\neq0$ for a time interval $t_1$, then at $q=q_2\neq0$ for $t_2$, followed by $q=q_1\neq0$
for another time $t_3$, and finally again $q=q_2\neq0$ for interval $t_4$.
For realistic solutions, $q_1$ and $q_2$ must be of the same sign, dictated by the element under consideration.
\subsection{HYPERFINE SPIN-1}
Here we directly show the analytic solution for $t_1$ and $t_2$
as functions of $q_1$ and $q_2$ for $f=1$. The state vector at time $t=t_1+t_2$ is given by
\bea
\left( \ba{c} u(t) \\ v(t) \ea \right)
&=&
\left( {\rm cos} \Omega_2 t_2 - i
\frac{ \vec H_{eff,2} \cdot \vec \tau}{\Omega_2} {\rm sin} \Omega_2 t_2 \right) \\
&\times& \left( {\rm cos} \Omega_1 t_1 - i
\frac{ \vec H_{eff,1} \cdot \vec \tau}{\Omega_1} {\rm sin} \Omega_1 t_1 \right)
\left( \ba{c} u(0) \\ v(0) \ea \right),\nonumber
 \eea
where $H_{eff,i}$ and $\Omega_i$ are defined as Eq.~(\ref{def-Omega}). We obtain the solution of $t_i$ for the equation $v(t_1+t_2)=0$ as
 \bea
 {t_1} &=&\pm \frac{1}{\Omega_1} \cot ^{-1}\left(\frac{|E_0-E_2|}{2\sqrt{3}\Omega_1} \left[
 {\frac{2q'_1+1}{2q'_2+1} C(q'_1,q'_2)  } \right]^{1/2}\right) \nonumber \\
 {t_2} &=&\mp \frac{1}{\Omega_2}\cot ^{-1}\left(\frac{|E_0-E_2|}{2\sqrt{3}\Omega_2} \left[
 {\frac{2q'_2+1}{2q'_1+1} C(q'_1,q'_2)  } \right]^{1/2}\right), \nonumber \\
 \label{sie}
 \eea
where $C(q'_1,q'_2)\equiv 2q'_2-3-6q'_1-12q'_1q'_2$.
For convenience we have defined $q'_i={q_i}/(E_0-E_2)$.
From Eq.~(\ref{sie}), given one solution $(t_1^*,t_2^*)$,
the other solutions can be obtained as  $(t_1^*+n\pi/\Omega_1,t_2^*+m\pi/\Omega_2)$ and
$(-t_1^*+n\pi/\Omega_1,-t_2^*+m\pi/\Omega_2)$, where $n$ and $m$ are integers.

\begin{figure}
\vspace{-0.0cm}
\includegraphics[width=0.49\textwidth]{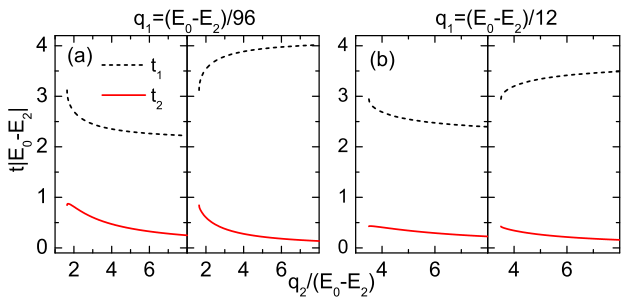}
\vspace{-0.6cm} \caption {In frame (a) and (b), plots
$t_1$ and $t_2$ as function of $q_2$ for two sets of solutions.
In the frame (a) and (b) the values of $q_1/(E_0-E_2)$ are
 $1/96$ and $1/12$ respectively.}
\label{fig3}
\end{figure}

Interestingly no solution is found when the condition
$(2q'_1+1)(2q'_2+1)(2q'_2-3-6q'_1-12q'_1q'_2)\geq0$  does not hold. The conditions on $q'_i$ for the existence of solutions can be classified into two cases according to signs of $q'_i$; case(1) corresponds to $q'_1,q'_2\leq0$, and case(2) corresponds to $q'_1,q'_2\geq0$. The conditions for both cases are listed below in more details.
 \bea
  \mbox{Case(1)}  &:&\texttt{\ \ \ \ }
 q'_1 \leq -\frac{1}{2} ,\texttt{\ \ \ }-\frac{1}{2}\leq q'_2\leq \frac{3+6q'_1}{2-12q'_1}\leq0
 \texttt{\ \ } \nonumber \\
 &\texttt{or}& \texttt{\ }
 -\frac{1}{2}\leq q'_1 \leq 0,\texttt{\ \ \ }q'_2 \leq -\frac{1}{2}.\nonumber \\
  \mbox{Case(2) }  &:&\texttt{\ \ \ }
 0\leq q'_1 \leq\frac{1}{6},\texttt{\ \ \ } \frac{3}{2}\leq\frac{3+6q'_1}{2-12q'_1}\leq q'_2.\nonumber
 \eea
Case$(1)$ applies to $^{23}$Na and case$(2)$ applies to $^{87}$Rb. When $q'_1=0$ the results reduce to that in Section~\ref{sec:one}. In both cases either $|q'_1|$ or $|q'_2|$ can not be too large, otherwise no solution is found. In Fig.~\ref{fig3}(a)(b) we show the solutions for $t_i$ with the choices $q'_1=1/96$ and $1/12$ respectively for case (2), and solutions exist only when $q'_2\geq 49/30$ and $7/2$ accordingly.

For the $^{87}$Rb case, by choosing $\tilde{U}a_B=(2\pi)30$ Hz, we have $E_0-E_2\approx264$ Hz. So the required magnetic fileds for $q'_1=1/96$ and $1/12$ are $B\approx 0.20$ G and $0.55$ G, respectively; for $q'_2=49/30$ and $7/2$, $B\approx 2.45$ G and $3.59$ G, respectively. Solutions of $(t_1,t_2)$ are $(11.8,3.2)$ ms and $(11.2,1.61)$ ms
for $(q'_1,q'_2)=(1/96,49/30)$ and $(1/12,7/2)$, respectively.

\subsection{HYPERFINE SPIN-2}
For $f=2$ spin systems in the non-zero magnetic field the time-evolution~(\ref{evolution}) operator is given as
    \bea
    \hat{U}(t_1,t_2,t_3,t_4)&=& e^{ {-i H_2 t_4}} e^{ {-i H_1 t_3}  }
    e^{ {-i H_2 t_2} }  e^{{-i H_1 t_1}  }.
    \eea
Here $H_i$ represents the Hamiltonian in Eq.\ (\ref{H2}). If $q_1=0$, the time-evolution operator is just Eq.\ (\ref{evolution}). By a similar method in section \ref{sec:two} the solutions $t_i$'s can be obtained. Here we also have limited ourselves to $t_{tot}=t_1+t_2+t_3+t_4<200$,
in units of $1/\varepsilon$. There is a symmetry
property: the solutions are the same when
$q_i=q_i^*$ and $\theta=\theta^*$ or $q_i=-q_i^*$ and $\theta=-\theta^*$.

\begin{figure}
\vspace{-0.0cm}
\includegraphics[width=0.48\textwidth]{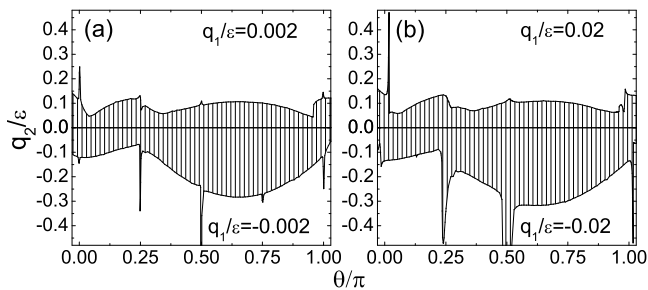}
\vspace{-0.6cm} \caption {In frame(a) and (b) the parameter regimes where
we can not find solutions are shown as shaded regions,
where the values of $q_1/\varepsilon$ are $0.002$
and $0.02$ respectively in the upper panels,
and the values of $q_1/\varepsilon$ are
$-0.002$ and $-0.02$ respectively in the lower panels.}
\label{fig4}
\end{figure}

In Fig.\ \ref{fig4}.(a)(b) no solution is found in the shaded regions, where the values of $q_1/\varepsilon$
are $0.002$ and $0.02$ for the upper-half of the panels, and the values of $q_1/\varepsilon$ are $-0.002$ and $-0.02$ for the lower-half of the panels. Compared with the result of $q_1=0$
(see Fig.\ \ref{fig2}.(a)) we find that when the absolute value of $q_1$ increases, the main change occurs in the region for negative $q$, where the total area with no solution found increases. Especially the peaks near $\theta=\pi/4$ and $\pi/2$ grow quickly.
Thus for $0 \le \theta \le \pi$ and $q < 0$
(and similarly for $\pi \le \theta \le 2 \pi$ and $q > 0$), $|q_1|$ cannot be too large in experiments.
Besides there are also some changes in the peak positions, and some peaks are disappeared.

For $^{23}$Na and $^{87}$Rb in $f=2$, if $q_1/\varepsilon=-0.002$ and $\varepsilon t_i<10$, then solutions exist when $q_2/\varepsilon\lesssim-0.21$.
For example with $(q_1/\varepsilon,q_2/\varepsilon)=(-0.002,-0.24)$
there is one solution of $\varepsilon \cdot t_{i=1-4} = (3.76, 4.56, 3.72, 2.38)$ for $^{23}$Na
and $(3.80, 4.38, 3.34, 2.49)$ for $^{87}$Rb.
By choosing $\tilde{U}a_B=(2\pi)30$ Hz, the $B$ fields corresponding to $(q_1,q_2)$ are $(0.20,2.28)$ G and $(0.24,2.7)$ G, and the solutions for $t_{i=1-4}$ are $(0.624, 0.757, 0.618 , 0.395)$ ms and $(1.75, 2.01, 1.54, 1.14)$ ms for Na and $^{87}$Rb, respectively.

\section{Conclusion} \label{sec:four}
In this paper we propose a method to  dynamically generate a two-particle total hyperfine spin-singlet. For hyperfine spin-1 systems, by allowing the system to evolve in two time intervals, with a specific magnetic field in each interval. For hyperfine spin-2 systems, a similar scheme with four time intervals can also achieve our goal. For some special cases, such as negative $q$ and $E_2 \approx E_0<E_4$, the singlet state is however hard to obtain unless long evolution times are used. Except in these very special circumstances,  our proposed scheme should be easy to implement in realistic experiments.  Preparation of these singlet pairs would allow us to create some exotic many-body states, and can also be useful in quantum information applications.
\acknowledgements
This research is supported by the  National Science Council of Taiwan.
\appendix
\section{Influence of inaccuracies in the time intervals used in evolution}\label{app1}
Here we provide some details on our error estimates on the final wavefunction. First we consider $f=1$.
With the condition $v(t_1,t_2)=0$, we can get the analytic form of
$|v(t_1+\Delta t_1,t_2+\Delta t_2)|^2\equiv |\Delta v|^2$ as
\bea
|\Delta v|^2 & \simeq & \frac{2}{9}[ (E_{20}\Delta t_1)^2 +(2q_2\Delta t_2)^2
\label{dv1}
\\
&+&(\Delta t_1\Delta t_2)(12q_1q_2+3E_{20}^2-6q_1E_{20}-2q_2E_{20}) ],\nonumber
\eea
where $E_{20}\equiv E_2-E_0$. Here we have assumed that $\Delta t_1$ and $\Delta t_2$
are small enough. In addition, Eq.\ (\ref{dv1}) can also be generalized to an inequality,
\bea
  |\Delta v|^2 &<&\left[  \Omega_1 | \Delta t_1| + \Omega_2 |\Delta t_2|\right]^2 \nonumber \\
  &\leq& \left[ \max(\Omega_1 ,\Omega_2 )(|\Delta t_1|+|\Delta t_2|)\right]^2\equiv\Lambda _1.
\label{dv2}
\eea
To get this inequality the condition, $(2q'_1+1)(2q'_2+1)(2q'_2-3-6q'_1-12q'_1q'_2)\geq0$,
has been used. In a special case where $q_1=0$, $ \Omega_1$ reduces to $|E_{20}|/2$ seen in section \ref{sec:one}.

This last formula for $\Lambda_1$ gives an upper limit on the error produced due to
the inaccuracies of time intervals employed in an experiment. If $\Lambda_1=0.02$, then
 $|u|$ (fidelity) would be higher than $0.99$.

For $f=2$ system we simply use the estimate $|\Delta v|^2 < \Lambda_2\equiv
[ \max(\Delta \omega_1,\Delta \omega_2)(|\Delta t_1
|+|\Delta t_2|+|\Delta t_3|+|\Delta t_4|) ]^2$, where
$\Delta \omega_1$ and $\Delta \omega_2$ are defined as the maximum
of $|\omega_{1,i}-\omega_{1,j}|/2$ and $|\omega_{2,i}-\omega_{2,j}|/2$, respectively.
Here $\omega_{1,i}$ and  $\omega_{2,i}$ represent the eigenvalues of the
Hamiltonian in Eq.~(\ref{H2}) at $q=q_1$ and $q=q_2$, respectively.
Although we have not obtained the analytic form for the $f=2$ systems,
we verified numerically that this formula for
$\Lambda_2$ works well in practice.  The fidelities discussed in text were estimated
using this formula for $\Lambda_2$.


\begin{thebibliography}{plain}
\bibitem{note} Though not all of the wells in their experiment are doubly occupied.
\bibitem{note1} We understand this as follows.  Consider $\theta \approx \frac{ 3 \pi}{2}$,
where $E_0 \approx E_2 > E_4$.  The state $| 4 0\rangle _F$ is well separated from the other two states
 at $q=0$.  At finite $q > 0 $, the diagonal parts of the quadratic Zeeman energy satisfy
 $\frac{17}{7} \sim 2 > \frac{4}{7}$, thus shifting the $|20\rangle _F$ and $ |0 0\rangle _F$ states
 further up and away from the lowest $| 4 0\rangle _F$ level.  The time evolution thus
 becomes inefficient to change the amplitude of our state in the $|4 0\rangle _F$ component.
% For large $q$, it is more convenient to think in terms of the state in terms of the $m$ basis.
% In this case, the differences in the interaction energies are inefficient to
% modify the state from the initial $|0,0>_m$ state.
Similar reasoning applies to $\theta = \frac{\pi}{2}$ but $q < 0$.
\bibitem{note2} We understand this as follows.  At $\theta = \frac{\pi}{4}$,
 $E_2$ and $E_4$ are equal.  The time evolution during the $q=0$ time
 intervals therefore does not modify the relative phase between the $|2 0\rangle _F$ and
 $|4 0\rangle _F$ amplitudes,  limiting the capability of the present scheme.  The other
 cases can be understood in a similar manner.
\end{thebibliography}
\end{document}